\documentclass[final,3p,sort&compress]{elsarticle}

\usepackage{
	amsmath,amssymb,
    color
}
\usepackage[colorlinks,linkcolor=blue,citecolor=blue,urlcolor=blue]{hyperref}
\hypersetup{pdfstartview={XYZ null null 1.15}}

\journal{Special Issue: Localisation2020}

\begin{document}

\begin{frontmatter}
	
	\title{Green's functions on a renormalized lattice: An improved method for the integer quantum Hall transition}
	
	\author[UR,MST]{Martin Puschmann}
	\author[MST]{Thomas Vojta}
	\address[UR]{Institute of Theoretical Physics, University of Regensburg, 93053 Regensburg,  Germany}
	\address[MST]{Department of Physics, Missouri University of Science and Technology, Rolla, Missouri 65409, USA}
	
	\begin{abstract}
		We introduce a performance-optimized method to simulate localization problems on bipartite tight-binding lattices. It combines
		an exact renormalization group step to reduce the sparseness of the original problem with the recursive Green's function method.
		We apply this framework to investigate the critical behavior of the integer quantum Hall transition of a tight-binding Hamiltonian
		defined on a simple square lattice. In addition, we employ an improved scaling analysis that includes two irrelevant exponents
		to characterize the shift of the critical energy as well as the corrections to the dimensionless Lyapunov exponent.
		We compare our findings with the results of a conventional implementation of the recursive Green's function method,
		and we put them into broader perspective in view of recent development in this field.
	\end{abstract}
	
	
\end{frontmatter}

\section{Introduction}
The integer quantum Hall (IQH) transition is a paradigmatic quantum phase transition in the realm of Anderson localization\ \cite{EveM08}. It appears when a two-dimensional electron gas is subject to a perpendicular magnetic field, leading to discrete and highly degenerate Landau levels\;(LLs) (within the framework of noninteracting electrons). Disorder lifts the degeneracy, and the LLs broaden into Landau bands\;(LBs). All electronic states are spatially localized but their localization length $\xi$  diverges at a critical energy $E_\mathrm{c}$ in the center of each Landau band. An IQH transition occurs if the Fermi level crosses one of these critical energies. Analogous to other Anderson localization transitions, the divergence of the localization length is usually described by a power-law dependence, $\xi(E)\sim|E-E_\text{c}|^{-\nu}$, on the energy $E$ where the critical localization length exponent $\nu$ is a universal and finite number.

Early numerical investigations of the IQH transition gave $\nu$ values in the range of 2.3 to 2.4\;\cite{HucK90,Huc95,CaiRR03} and agreed well with experimental results\;\cite{LiCT05,LiVX09}. Roughly a decade ago, however, Slevin and Ohtsuki\;\cite{SleO09} performed a careful scaling analysis of the Lyapunov exponents of the semi-classical Chalker-Coddington\;(CC) network model\;\cite{ChaC88}. Their result, $\nu=2.593\;[2.578,\;2.598]$ (where the numbers in brackets mark the confidence interval), was clearly higher than what was observed in earlier numerical investigations as well as experiments.
In the following years, several other investigations confirmed this new, larger value of $\nu$\;\cite{AmaMS11,SleO12,ChaGKS20,FulHA11} but more recently, the reported results have been more diverse. A value of $\nu=2.37(2)$ (where the number in brackets gives the error of the last digit) was obtained for a structurally disordered CC network\;\cite{GruKN17,KluNS19}. A similar value, i.e. $\nu=2.33(3)$, was reported for a Dirac fermion model\;\cite{SbiDMG20arXiv}. Zhu \textit{et al.}\;\cite{ZhuWBW19} analyzed the scaling behavior of conducting channels and obtained $\nu=2.48(2)$, a value different from most other investigations. As all of these models are supposed to feature the same universal critical behavior, the value of the localization length exponent of the IQH transition must be considered an open problem.

Conventionally, the discrepancies between different theoretical values have been attributed to strong, long-ranged finite-size effects which either lead to power-law corrections to scaling governed by an irrelevant exponent $y$ or even to
marginal (logarithmic) corrections. Almost all investigations have shown that $y$ is small and thus difficult to analyze. Literature values range from $y=0.15$ to $0.7$.
Differences between theoretical and experimental values may be caused by Coulomb interactions that exist in experiments but are neglected in numerical simulations based on single-particle models.
Recently, Zirnbauer\;\cite{Zirnbauer19} proposed a different scenario. Based on a conformal field theory for the IQH transition, he proposed that all scaling is of marginal type, implying that asymptotically $\nu\rightarrow\infty$ and $y\rightarrow 0$. Finite non-zero values for $\nu$ and $y$ would then just represent effective exponents that depend on the model and the distance from criticality at which the data are taken. Verifying or falsifying this theory numerically is extremely difficult because the identification of logarithmic scaling requires enormous system sizes. Even marginal corrections to conventional power-law scaling are not easy to identify because accurate data are needed to discriminate between power-law behavior with a small $y$ and logarithmic behavior.

As many of the high numerical values for the localization length exponent ($\nu \approx 2.6$) stem from the CC network model,
Gruzberg et al.\;\cite{GruKN17} suggested that the conventional CC network is, perhaps, too regular and does not contain all types of disorder relevant at the IQH transition. Indeed, for a structurally disordered CC network, they obtained the above-mentioned lower value of $\nu=2.37(2)$. It is therefore important to determine whether or not the paradigmatic CC network correctly captures the physics of disordered noninteracting electrons close to the IQH transition.
This motivated us to analyze the IQH transition in a microscopic tight-binding model of noninteracting electrons on a square lattice. In Refs.\ \cite{PusCSV19} and \cite{PusCSV20arXiv}, we studied the transitions in the lowest LB in cylinder and strip geometries, respectively, giving us access to the behavior of the states both in the bulk and at the edges of the system.
Interestingly, our observed value $\nu=2.58(3)$ is in agreement with the results for the (conventional) CC network.
As our scaling analysis was limited to larger system sizes for which the finite-size effects are weaker, we were unable to conclusively discriminate between power-law and logarithmic corrections to scaling.

The purpose of the present paper is twofold. First, we combine two established numerical methods to increase the performance of our calculation. Specifically, we employ a numerically exact renormalization-group step\;\cite{Aok80,Aok82,MonG09} to obtain a
renormalized tight-binding model with half the original number of lattice sites. We then use the recursive Green's function approach to calculate critical properties based on the renormalized system. This approach improves the computational performance by a factor of four compared to a recursive Green's function calculation on the original lattice, as employed in our earlier works \cite{PusCSV19,PusCSV20arXiv}. We use the performance gain to increase the system size by a factor of two and to improve the accuracy of our data. Second, we employ a modified version of the finite-size scaling analysis. It is motivated by the observation that irrelevant corrections to the critical energy appear to be governed by a much larger irrelevant exponent $y'$ than the corrections to the dimensionless Lyapunov exponent which are governed by the leading irrelevant exponent $y$. Including both $y$ and $y'$
in the scaling analysis, we obtain $\nu=2.60(2)$, $y=0.31(3)$, $y'=1.4(2)$, and a critical dimensionless Lyapunov exponent of $\Gamma_\mathrm{c}=0.806(6)$.

The rest of the paper is organized as follows. In Sec.\ \ref{sec:model}, we introduce the tight-binding model and describe
the renormalization of the lattice. We also discuss the recursive Green's function method and the details of our sophisticated scaling ansatz. Section\ \ref{sec:results} first reports the results of a heuristic finite-size scaling analysis which motivates the sophisticated compact scaling approach, whose results are presented afterwards. We conclude in Sec.\ \ref{sec:summary}.

\section{Model \& Methods}\label{sec:model}
\subsection{Tight-binding model}
We consider a tight-binding model of non-interacting electrons defined on a regular square lattice of lattice constant $l$. In site representation, the Hamiltonian reads
\begin{align}
	\mathbf{H}=\sum_j u_j |j\rangle\!\langle j| + \sum_{\langle j,k \rangle} t_{jk} |j\rangle\!\langle k| \label{eq:hamiltonian_general}
\end{align}
where $|j\rangle$ denotes the Wannier state at site $j$. Disorder is represented via independent random potentials $u_j$ drawn from a uniform (box) distribution on the interval $\left[-W/2,\ W/2\right]$, where $W$ characterizes the disorder strength. The hopping matrix elements (bonds) $t_{jk}$
connect nearest neighbor sites only; they have a constant magnitude which we set to unity.

To study the quantum Hall physics, we apply a uniform magnetic field perpendicular to the plane of the lattice. Assuming that lattice lies in the $xy$-plane (with the nearest neighbor bonds parallel to the coordinate axes), the field reads $\mathbf{B}=B\hat{z}$, and the vector potential (in Landau gauge) is given by $\mathbf{A}=(0,Bx,0)$. The leading effect of the field on the Hamiltonian (\ref{eq:hamiltonian_general}) is the creation the Peierls phases\;\cite{Pei33}
\begin{equation}
	\varphi_{jk}=\frac{\mathrm{e}}{h}\int_{j}^{k} \mathbf{A}\cdot \mathrm{d}\mathbf{r}  = 2\pi \frac{eB}{\hbar}\left(\frac{x_j+x_k}{2}\right)\left(y_k-y_j\right)\quad,
	\label{eq:Peierls}
\end{equation}
describing the phase change of an electronic wave function when hopping between two neighboring sites $j$ and $k$. Here, $x_j$ and $y_j$ are the coordinates of lattice site $j$. In the presence of the magnetic field, the hopping matrix elements thus read $t_{jk}=\exp(i\varphi_{jk})$.
According to eq.\ (\ref{eq:Peierls}), the magnetic field has no effect, $\varphi_{ij}=0$, for hopping in the $x$ direction whereas $\varphi_{ij}=\pm 2\pi\Phi x_i$ for hopping in the $y$ direction. Here, and $\Phi=Bl^2e/h$ is the dimensionless magnetic flux through a unit cell of area $l^2$. If one wishes to implement periodic boundary conditions for the Hamiltonian (\ref{eq:hamiltonian_general}), the Peierls phases lead to constraints for
the possible values of the flux $\Phi$. Actually, boundary conditions $y$ direction do not constrain $\Phi$, but the number of sites $L_\mathrm{x}$ in the $x$ direction must be an integer multiple of $1/\Phi$. For quasi-one-dimensional lattices in strip or cylinder geometry, it is thus convenient to choose the $x$ direction for the long (quasi-infinite) side of the lattice and the $y$ direction for the short side.
The Hamiltonian (\ref{eq:hamiltonian_general}) for such a quasi-one-dimensional lattice now takes the block-tridiagonal form
\begin{equation}
	\mathbf{H}=\begin{pmatrix}
		\mathbf{H}_{1} & \mathbf{I} &  &  &  \\
		\mathbf{I} & \mathbf{H}_{2} & \mathbf{I} & & \\
		& \mathbf{I}  & \mathbf{H}_{3} & \ddots & \\
		&  & \ddots & \ddots & \mathbf{I}\\
		&  &  & \mathbf{I} &
		\mathbf{H}_{N}
	\end{pmatrix}
	\quad\text{with}\quad\mathbf{H}_x=\begin{pmatrix}
		u_{x,1} & \mathrm{e}^{i \varphi_x} &  &  &  \mathrm{e}^{-i \varphi_x}\\
		\mathrm{e}^{-i \varphi_x} & u_{x,2} & \mathrm{e}^{i \varphi_x} & & \\
		& \mathrm{e}^{-i \varphi_x}  & u_{x,3} & \ddots & \\
		&  & \ddots & \ddots & \mathrm{e}^{i \varphi_x}\\
		\mathrm{e}^{i \varphi_x} &  &  & \mathrm{e}^{-i \varphi_x} &
		u_{x,L}
	\end{pmatrix} \; .\label{eq:Hamiltonian_block}
\end{equation}
Here, $\varphi_x= 2 \pi \Phi x$. The layer matrices $\mathbf{H}_{x}$ contain the potentials and hopping matrix elements within the layer at position $x$. The interlayer connections take a particularly simple form, they are represented by identity matrices $\mathbf{I}$.

\subsection{Renormalized lattice}

We now focus on mapping the Hamiltonian (\ref{eq:hamiltonian_general}) onto a new, renormalized Hamiltonian with a smaller number of lattice sites. For this purpose, we employ Aoki renormalization group steps \cite{Aok80,Aok82,MonG09}. Each step consists of eliminating a single site and renormalizing the potentials and hopping matrix elements of the remaining sites. Specifically, if lattice site $k$ is eliminated, the potentials and hopping matrix elements of all sites connected to it are renormalized as
\begin{align}
	t'_{ij} &= t_{ij}+\frac{t_{ik} t_{kj}}{E-u_k}\;, \label{eq:tprime}\\
	u'_{i}  &= u_i+ \frac{t_{ik} t_{ki}}{E-u_k} \label{eq:uprime}
\end{align}
where $E$ is the energy of the state in question. In principle, these steps can be iterated ad infinitum until all sites have been eliminated,
providing an exact solution to the problem. However, in this process the renormalized hopping matrix elements proliferate and become more and more
long-ranged. This strategy this therefore not effective numerically.
We will instead use the Aoki renormalization group steps as ``preconditioner'' or accelerator for subsequent calculations, specifically for the recursive Green's function approach. To this end, we eliminate every second site of the original lattice arriving at a renormalized square lattice
with half the number of sites.

\begin{figure*}
	\centering
	\includegraphics{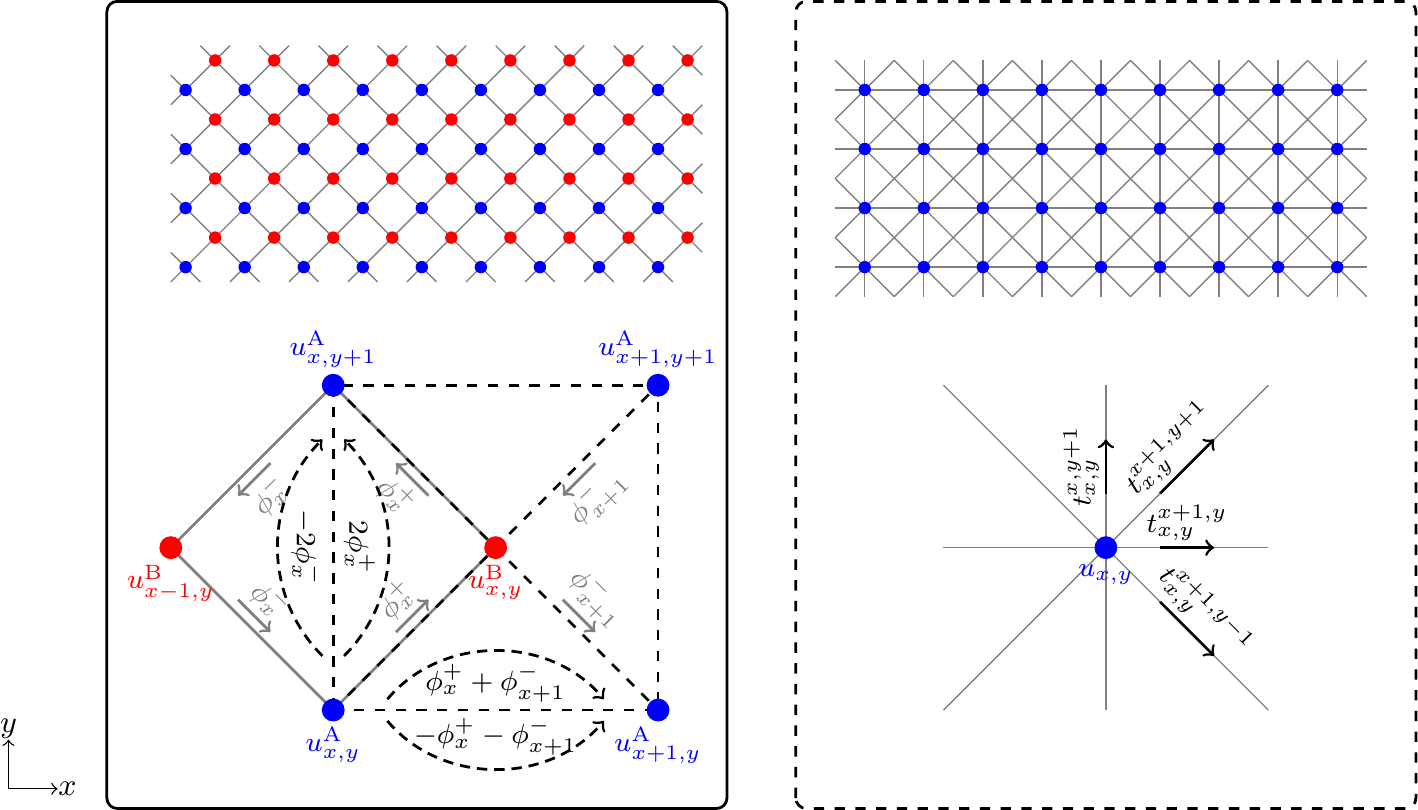}
	\caption{Renormalization group framework applied to the simple square lattice (left panel), leading to a simple square lattices with additional diagonal bonds (right panel) but half the number of lattice sites. Here all $B$-type lattice sites (red dots) have been eliminated, leading to bonds between $A$-type sites (blue dots) only. Each panel shows the lattice (upper part) and a unit cell (lower part). The lower left panel sketches how the Peierls phases of the $A\!-\!B$ bonds of the simple square lattice combine to phase changes (black dashed) of effective  $A\!-\!A'$ bonds in the renormalized system.  }
	\label{fig:lattice}
\end{figure*}

In detail, the renormalization process works as follows. In the square lattice (or any bipartite lattice) we can form two disjoint sublattices, $A$ and $B$, such that each site on the $A$ sublattice has bonds to $B$ sites only, and vice versa. This allows us to perform the Aoki renormalization group step for all the $B$ sites independently from each other. This introduces new effective $A-A'$ bonds between nearest and next-nearest neighbors on the
A sublattice. In this way, we reduce the number of lattice sites by two. We will see later that the existence of next-nearest neighbor bonds does not significantly affect the performance of the recursive Green's function method. We also note that the renormalized lattice is not bipartite implying that the procedure cannot be iterated without creating even longer bonds.
Figure\ \ref{fig:lattice} illustrates the procedure as applied to a simple square lattice; the renormalized system is a square lattice with additional diagonal bonds.

We now apply this framework to the IQH Hamiltonian\ (\ref{eq:hamiltonian_general}). To keep the later equations simple, we rotate the original lattice by $\pi/4$ and scale the lattice constant by $1/\sqrt{2}$ such that its next-nearest-neighbor bonds (which will become the nearest-neighbor bonds of the renormalized lattice) have unit length. This is compensated by an additional factor $2$ in the Peierls phases
\begin{align}
	\varphi_{jk}=4\pi\Phi\left(\frac{x_j+x_k}{2}\right)\left(y_k-y_j\right)\quad.
\end{align}
such that the flux $2\phi_x^+ + 2\phi_x^-=2\pi\Phi$ through a unit cell is constant and equal to that of the original Hamiltonian (\ref{eq:hamiltonian_general}). Here, $\phi_x^\pm=\pm 2\pi \Phi(x\pm 1/4)$ are the Peierls phases of the links between two nearest-neighbor sites, as illustrated in the left panel of Fig.\ \ref{fig:lattice}. Moreover, anticipating the renormalization, we relabel the sites of the original lattice in terms of the sublattice label and the $x$ and $y$ positions of the (surviving) A sites. Correspondingly, the potentials are labeled as $u^A_{x,y}$ and $u^B_{x,y}$.

We now eliminate all $B$ sites via Aoki renormalization group steps (\ref{eq:tprime}) and (\ref{eq:uprime}). This leads to the renormalized Hamiltonian defined on the sites of the $A$ sublattice only,
\begin{align}
	\mathbf{H}_\mathrm{ren}=\sum_{x,y} u_{x,y} |x,y\rangle\!\langle x,y| + \sum_{x,y,x',y'} t_{x,y}^{x',y'} |x,y\rangle\!\langle x',y'| \label{eq:Hamiltonian_red}
\end{align}
with effective potentials and hopping matrix elements
\begin{align}
	\begin{aligned}
		u_{x,y} &=u^A_{x,y}+\kappa_{x,y}+\kappa_{x-1,y}+\kappa_{x-1,y-1}+\kappa_{x,y-1} \\
		t_{x,y}^{x+1,y} &= e^{-i\chi}\kappa_{x,y} + e^{i\chi}\kappa_{x,y-1}\\
		t_{x,y}^{x,y+1} &= e^{i\varphi_x+i\chi}\kappa_{x,y} + e^{i\varphi_x-i\chi}\kappa_{x-1,y} \\
		t_{x,y}^{x+1,y+1} &= e^{i\varphi_x+2i\chi}\kappa_{x,y}\\
		t_{x,y}^{x+1,y-1} &= e^{-i\varphi_x-2i\chi}\kappa_{x,y-1}
	\end{aligned}
\end{align}
where $\varphi_x=4\pi\Phi x$, $\chi=\pi\Phi$, and $\kappa_{x,y} = 1/(E- u^B_{x,y})$. The effective potentials $u_{x,y}$ are renormalized by contributions based on $A\!-\!B\!-\!A$ links beginning and ending at the same $A$ site. Since forward and backward hopping cancels the phase changes, the additional contributions to $u_{x,y}$ are real and given by the sum of inverse reduced potentials $\kappa$ of all four neighboring $B$ sites.
The new hopping matrix elements  $t_{x,y}^{x',y'}=(t_{x',y'}^{x,y})^\ast$ between $A$ sites $(x,y)$ and $(x',y')$ stem from $A\!-\!B\!-\!A'$ links
with either one or two possible connection paths. The left panel of Fig.\ \ref{fig:lattice} sketches the accumulated phase change along the subpaths ${A\!-\!B\!}$ and $\!B\!-\!A'$. The magnitudes of the renormalized hopping matrix elements (for each path) are given by the inverse reduced potential $\kappa$ of the intermediate $B$ sites.

As a result, the renormalized tight-binding Hamiltonian is more complex than the original one because it contains next-nearest neighbor couplings, and the hopping matrix elements have nonuniform magnitude.  As the Peierls phases still depend on $x$ only, quasi-one-dimensional systems
with periodic boundary conditions in $y$-direction can be easily constructed. In this geometry, the renormalized Hamiltonian again takes a block-tridiagonal form,
\begin{align}
	\mathbf{H}_\mathrm{ren}&=\begin{pmatrix}
		\mathbf{H}_{1} & \mathbf{V}_{1} &  &  &  \\
		\mathbf{V}_{1}^\dagger & \mathbf{H}_{2} & \mathbf{V}_{2} & & \\
		& \mathbf{V}_{2}^\dagger & \mathbf{H}_{3} & \ddots & \\
		&  & \ddots & \ddots & \mathbf{V}_{N-1}\\
		&  &  & \mathbf{V}_{N-1}^\dagger &
		\mathbf{H}_{N}
	\end{pmatrix}
	\label{eq:H_ren_block}
\end{align}
where both the diagonal and the off-diagonal blocks have tridiagonal forms as well,
\begin{align}
	\mathbf{H}_x&=\begin{pmatrix}
		u_{x,1} & t_{x,1}^{x,2} &  &  &  t_{x,1}^{x,L}\\
		t_{x,2}^{x,1}  & u_{x,2} & t_{x,2}^{x,3} & & \\
		& t_{x,3}^{x,2} & u_{x,3} & \ddots & \\
		&  & \ddots & \ddots & t_{x,L-1}^{x,L}\\
		t_{x,L-1}^{x,1} &  &  & t_{x,L}^{x,L-1} &
		u_{x,L}
	\end{pmatrix}~,\quad\mathbf{V}_x=\begin{pmatrix}
		t_{x,1}^{x+1,1} & t_{x,1}^{x+1,2} &  &  &  t_{x,1}^{x+1,L}\\
		t_{x,2}^{x+1,1}  & t_{x,2}^{x+1,2} & t_{x,2}^{x+1,3} & & \\
		& t_{x,3}^{x+1,2} & t_{x,3}^{x+1,3} & \ddots & \\
		&  & \ddots & \ddots & t_{x,L-1}^{x+1,L}\\
		t_{x,L}^{x+1,1} &  &  & t_{x,L}^{x+1,L-1} & t_{x,L}^{x+1,L}
	\end{pmatrix}\;.\label{eq:Hamiltonian_red_block}
\end{align}
Note that the interlayer coupling matrices $\mathbf{V}_x$ are not identity matrices anymore, but they are still sparse.

\subsection{Recursive Green's function approach}

We now consider quasi-one-dimensional systems defined on long cylinders (strips with periodic boundary conditions in the ``short'' direction) of circumference $L$ and length $N$ with $N \gg L$. Within the recursive Green's function algorithm, we determine the time-independent Green's function at energy $E$, defined as $\mathbf{G}(E)=\lim_{\eta\rightarrow 0} \left[(E+i\eta)\mathbf{I}-\mathbf{H}_\mathrm{red}\right]^{-1}$, where $\mathbf{I}$ is the identity matrix. For a quasi-one-dimensional system, the smallest positive Lyapunov exponent (inverse localization length) $\gamma$ can be computed from the Green's function between the first and last layers,
\begin{eqnarray}
	\gamma(E,L)=\lim\limits_{N\rightarrow \infty}\frac{1}{2N} \ln|\mathbf{G}^{N}_{1N}|^2\;,\label{eq:lyapunov}
\end{eqnarray}
Effectively, this expression measures the average exponential decay of the wave function between these two layers at energy $E$.
(Note that the upper index of $\mathbf{G}$ denotes the total number of layers in the system. The lower indices are the arguments, i.e.,
they mark between which layers the green's function is taken.)

Due to the block-tridiagonal structure of both the original and the renormalized Hamiltonians, the Green's function $\mathbf{G}^{N}_{1N}$ between the first and last layers for either system can be calculated efficiently in a recursive manner. In the following, we briefly outline this method on the example of the renormalized Hamiltonian (\ref{eq:Hamiltonian_red_block}); a more detailed description of the algorithm is given in Refs.\;\cite{Mac80,MacK83,KraSM84,SchKM84,Mac85,SchKM85,Huc95}.
We start with the Hamiltonian of an isolated single layer, $\mathbf{H}_{1}$. At each iteration step we add another layer $\mathbf{H}_{i}$ to the already existing layer stack. The coupling blocks $\mathbf{V}_{i-1}$ and $\mathbf{V}_{i-1}^\dagger$ are treated as perturbations to the stack consisting of $i-1$ layers. Using the Dyson equation, $\mathbf{G}^{N}_{1N}$ can be obtained by means of two recursion steps
\begin{align}
	\begin{aligned}
		\mathbf{G}^{N+1}_{1N+1} =&\mathbf{G}^{N}_{1N} \cdot \mathbf{V}_{N} \cdot \mathbf{G}^{N+1}_{N+1,N+1}\;, \\
		\mathbf{G}^{N+1}_{N+1,N+1} =& \left[(E+i\eta)\mathbf{I}-\mathbf{H}_{N+1}-\mathbf{\Sigma}_{N+1} \right]^{-1}\;.	
	\end{aligned}
\end{align}
Here, the self-energy $\mathbf{\Sigma}_{N+1} = \mathbf{V}_{N}^{\dagger}\cdot\mathbf{G}^{N}_{N,N}\cdot\mathbf{V}_{N}$ comprises the perturbation of the stack by the last layer. The iteration is initialized by $\mathbf{G}^1_{11}=\left[(E+i\eta)\mathbf{I}-\mathbf{H}_1\right]^{-1}$.

In the following, we make some remarks on the numerical implementation of the method. First, the matrix elements of $\mathbf{G}^{N}_{1N}$ typically decay very quickly with increasing $N$. To keep the numbers within the range that can be represented on the computer, we frequently (after every $\kappa=100$ iterations) extract the leading magnitude of $\mathbf{G}^{N}_{1N}$; mathematically, we replace the logarithm in eq.\ (\ref{eq:lyapunov}) by
\begin{equation}
	\ln|\mathbf{G}^{N}_{1N}|^2 = \ln |\mathbf{G}^{\kappa}_{1,\kappa}|^2 +\sum_{b=2}^{N/\kappa} \ln \frac{|\mathbf{G}^{b\kappa}_{1,b\kappa}|^2}{ |\mathbf{G}^{(b-1)\kappa}_{1,(b-1)\kappa}|^2}\;.
\end{equation}
Because $\ln|\mathbf{G}^{N}_{1N}|^2$ is a self-averaging quantity, one could perform the disorder average by simulating one very long sample, and the
statistical errors could be suppressed by simply making the sample longer while keeping the width fixed. For practical reasons including reasonable
run time per sample and the ease of parallelizing the calculation, it is more suitable to consider several long (but not extremely long) samples. We therefore approximate the limit ${N\rightarrow \infty}$ in eq.\ (\ref{eq:lyapunov}) by the finite length $N=10^6$ which is still three orders of magnitude large than our largest width $L$. Moreover, to reduce the effects of the open boundary at the beginning of the sample (layer 1), we perform  $3L$ iterations before we start measuring  $\gamma(E,L)$. Thus, $\gamma(E,L)=\frac{1}{2N} \ln|\mathbf{G}^{3L+N}_{3L,3L+N}|^2$. (We scale the number of discarded layers with $L$ because the localization length $\gamma^{-1}(E_\mathrm{c},L)\propto L$ at criticality.
For further data analysis and finite-size scaling, we focus on the dimensionless Lyapunov exponent $\Gamma(E,L)=L \langle\gamma(E,L)\rangle$. Here, $\langle\dots\rangle$ represents the average over a set of system realizations, and we use the standard deviation to describe the numerical accuracy of $\Gamma(E,L)$.

Let us also comment on the performance gain due to the Aoki renormalization steps before the recursive Green's function method.
The numerical effort of the recursive Green's function method scales as $L^3 N$ with the size of the system. The factor $N$
stems from the number of iterations, the factor $L^3$ is due to the inversion of dense $L\times L$ matrices. The Aoki renormalization steps rescale the lattice constant by a factor $\sqrt{2}$ in both directions. The numerical effort is thus reduced by a factor 4. One might be worried by the extra effort caused by the fact that the coupling matrices $\mathbf{V}_x$
of the renormalized Hamiltonian (\ref{eq:H_ren_block}) are not identity matrices, in contrast to the original
Hamiltonian (\ref{eq:Hamiltonian_block}). However, we employ sparse-matrix routines to implement the matrix-matrix multiplications involving $\mathbf{V}$. For $L\gg 1$, the numerical effort for these multiplications is negligible in comparison to the matrix inversion. Thus, the conceptual advantage of the original Hamiltonian, viz. having identity coupling matrices, does not significantly affect the performance.

The transfer-matrix technique is an alternative method to calculate the Lyapunov exponent $\gamma$ for quasi-one-dimensional systems. We have compared the results of the recursive Green's function and transfer matrix methods for the original Hamiltonian (without the Aoki renormalization) for sizes up to $L=256$ and found them to agree. The transfer-matrix technique is numerically less effective for the renormalized Hamiltonian due to the more complicated structure of its inter-layer coupling matrices $\mathbf{V}$. The recursive Green's function method also has the advantage that it can be easily applied to systems with
nontrivial geometry, ranging from  pyrochlore networks\;\cite{PusCS15} to randomly connected or percolating systems\;\cite{PusCHV20}.

\subsection{Sophisticated scaling approach}\label{sec:SSA}

To extract the critical energy as well as the critical exponents from the numerical data, we employ a finite-size scaling approach. It allows us to determine all critical parameters together via a combined fit of the dimensionless Lyapunov exponent data with a sophisticated scaling function,
\begin{equation}
	\Gamma(E,L)= F(x_\text{r}L^{1/\nu},\ x_\text{i,1}L^{-y_1},\ x_\text{i,2}L^{-y_2},\dots)~.
	\label{eq:fss_ansatz}
\end{equation}
It is expressed in terms of one relevant field $x_\text{r}L^{1/\nu}$ and one or more irrelevant fields $x_{\text{i},j}L^{-y_j}$. Here, $\nu$ and $y_j$ are positive numbers. The irrelevant fields decay with increasing system size $L$ and thus provide corrections to the leading power-law scaling governed by $\nu$. The scaling variables $x_r$  and $x_{i,\dots}$ depend, potentially in a nonlinear way, on the dimensionless distance from the critical point, $\rho=E/E_\mathrm{c}-1$. In the absence of additional information, one needs to assume the most general form for $F$. Expanding the  scaling function in terms of the scaling fields yields
\begin{eqnarray}
	F(x_{\text{r}} L^{1/\nu}, x_{\text{i}} L^{-y})=\sum\limits_{i=0}^{n_{\text{i}}}\sum\limits_{j=0}^{n_{\text{r}}}a_{ij}x_{\text{i}}^{i}x_{\text{r}}^{j}L^{j /\nu-iy}\quad.\label{eq:fss_general}
\end{eqnarray}
if only a single (leading) irrelevant scaling field is included. In a similar manner, we can expand the relevant and irrelevant scaling variables
\begin{eqnarray}
	x_{\text{r}}=\rho+\sum\limits_{k=2}^{m_{\text{r}}}b_{k}\rho^k\quad\text{and}\quad x_{\text{i}}=1+\sum\limits_{l=1}^{m_{\text{i}}}c_{l}\rho^l \label{eq:FSS_variables}
\end{eqnarray}
in terms of the distance from criticality $\rho$. The function $F$ is now fitted to the numerical data for $\Gamma(E, L)$, with
the critical energy $E_c$, the exponents $\nu$ and $y$, as well as the expansion coefficients $a_{ij}$, $b_k$ and $c_l$
serving as fit parameters. This kind of analysis is commonly used in the realm of Anderson localization transitions\;\cite{SleO99a}.

In contrast to the usual Anderson localization transitions which separate a localized phase from a delocalized phase, the IQH transition separates two localized phases. Due to this symmetry, we assume that only even powers of the relevant field  $x_\text{r}L^{1/\nu}$ contribute to the scaling function $F$. This is a nontrivial assumption because the numerical Lyapunov exponent data deviate from a perfect symmetry with respect to $E_\mathrm{c}$ for larger distances from criticality. To account for that, we keep even and odd terms in the expansion (\ref{eq:FSS_variables}) of $x_{\text{r}}$ in terms of the distance from criticality $\rho$. We do not make symmetry assumptions about the dependence of $F$ on the irrelevant fields  $x_{\text{i},j}L^{-y_j}$.

Before applying the sophisticated scaling approach to our numerical data, we perform a preliminary heuristic finite-size scaling analysis of our numerical data. It will be described in Sec.\ \ref{subsec:heuristic}. This analysis shows that the energy $E_\mathrm{min}(L)$, that marks the
position of the minimum of the dimensionless Lyapunov exponent $\Gamma(E,L)$ w.r.t.\ $E$, converges much faster towards its infinite-system limit than the value $\Gamma_\mathrm{min}(L)$ at this minimum. This suggests that we need two irrelevant exponents to describe the data over a wider range of system sizes, leading to the scaling function
\begin{eqnarray}
	F(x_{\text{r}}^2 L^{2/\nu},\ x_{\text{i}} L^{-y},\ x'_{\text{i}} L^{-y'}) =\sum\limits_{i=0}^{n_{\text{i}}} \sum\limits_{i'=0}^{n'_{\text{i}}}  \sum\limits_{j=0}^{n_{\text{r}}} a_{ii'j} x_{\text{i}}^{i} {x'}_{\text{i}}^{i'}x_{\text{r}}^{2j} L^{2j /\nu-iy-i'y'} \quad.\label{eq:fss}
\end{eqnarray}
Here, the exponent $y$ describes the leading irrelevant corrections, and $y'>y$ characterizes the subleading behavior (which should be visible in small systems only). Accounting, within this scaling function, for the asymmetries observed in the data for larger distances from criticality requires
higher order terms in the expansions (\ref{eq:FSS_variables}) of the scaling variables. Unfortunately, the number of fit parameters increases rapidly when higher expansion orders are included, leading to unstable fits. (This may explain why higher order terms in the expansions of the irrelevant scaling variables $x_\text{i}$ are rarely used in the literature.)

To overcome these difficulties, we restrict the expansion of each of the irrelevant scaling variables to its leading term, $x_\text{i}=x'_\text{i}=1$ but add a correction to the definition of the relevant scaling variable,
\begin{align}
	x_\text{r}= \rho+\sum\limits_{k=2}^{m_{\text{r}}}b_{k}\rho^k\quad\text{where}\quad \rho=\left(E-E_\text{c}+\sum\limits_{l=1}^{m_i} d_{l} L^{-ly'}\right)/E_\text{c}\quad.\label{eq:complex_scaling_field}
\end{align}
This means, we model the asymmetries by a shift in energy that decays with $L^{-y'}$. Note that this shift does not lead to a more general scaling function than eq.\ (\ref{eq:fss}) with unspecified scaling variables (\ref{eq:FSS_variables}), it just re-parameterizes the expansion terms.
The two irrelevant exponents $y$ and $y'$ now serve different purposes (and appear in different places in $F$) which stabilizes the fit and promotes an easier physical interpretation.

The expansion orders $n_\text{r}$, $n_\text{i}$, $n'_\text{i}$, $m_\text{r}$, and $m_\text{i}$ define the number $N_\text{P}$ of fit parameter are chosen such that we obtain fits of good quality. The least-square fits correspond to minimizing the cost function $\chi^2=\sum_i^{N_\text{D}} (\Gamma_i-F(E_i,L_i))^2/\sigma_i^2$, where the sum runs over the $N_\text{D}$ data points included in the fit. Each data point has energy $E_i$, system size $L_i$, and Lyapunov exponent value $\Gamma_i$ with a standard deviation $\sigma_i$. To analyze the fit quality, we consider the (reduced) mean-squared deviation $\tilde{\chi}^2= \chi^2/N_\text{F}$ and even more importantly the fit significance
\begin{align}
	p(\chi^2,N_\text{F})=\frac{\int_{\chi^2/2}^{\infty} t^{N_\text{F}/2-1} \exp(-t)\mathrm{d} t}{\int_{0}^{\infty} t^{N_\text{F}/2-1} \exp(-t)\mathrm{d} t}\quad,
\end{align}
where $N_\text{F}=N_\text{D}-N_\text{P}$ is number of free (fit) parameters. In the forthcoming analysis, our error estimates denote the standard deviation of the associated quantity.  

\section{Results} \label{sec:results}

Even in the clean case ($W=0$), the Hamilton\;(\ref{eq:hamiltonian_general}) leads to interesting physics. The interplay of the Peierls phases and the lattice periodicity leads to the so-called Hofstadter's butterfly\;\cite{Hof76,Lutt51}, describing the eigenvalue spectrum as a function of $E$ and $\Phi$. Hofstadter's butterfly features a fractal structure and symmetries with respect to $E=0$ and $\Phi=0.5$. In contrast to the case of free electrons, each Landau level has a nonzero intrinsic width $\delta$. The free-electron case is recovered for $\Phi\rightarrow 0$, where the intrinsic LL level width vanishes, $\delta\rightarrow 0$, the LL energies scale linearly with $\Phi$, and consecutive LLs are equidistant. Associated with the magnetic field is a characteristic length scale $L_\Phi\propto\sqrt{1/\Phi}$.

\begin{figure*}[t]
	\includegraphics{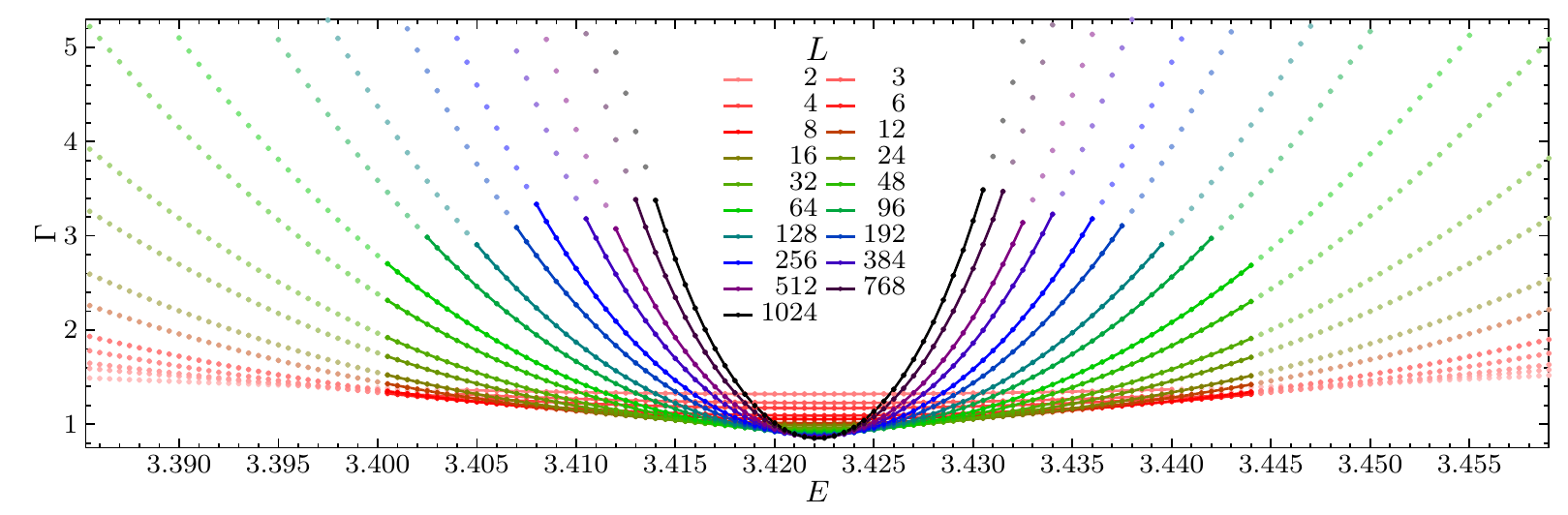}
	\caption{Dimensionless Lyapunov exponent $\Gamma(E,L)$ for $\Phi=1/10$ and several $L$ as a function of $E$ in vicinity of the IQH transition in the lowest LB. The statistical errors are well below the symbol size. The colored lines are guide to the eye, marking the data considered in the heuristic and the sophisticated scaling analysis (unused data is presented by isolated dots). Note that the plot is centered with respect to $E_\mathrm{c}=3.42215$.}
	\label{fig:transition}
\end{figure*}

In our previous work \cite{PusCSV19}, we analyzed the scaling properties of the IQH transition for the original, unrenormalized Hamiltonian\ (\ref{eq:hamiltonian_general}). We found that a flux of $\Phi=1/10$ provides the best compromise between competing optimization criteria. In the present study, we therefore focus on $\Phi=1/10$, $W=0.5$ and consider the energy-driven IQH transition in the lowest LB. We employ the renormalized Hamiltonian to calculate a large number of data points $\Gamma(E,L)$ in the vicinity of the critical energy $E_\mathrm{c}$.  Figure\;\ref{fig:transition} visualizes the resulting raw data for the dimensionless Lyapunov exponent. Each data point result from an average over $384$ strip realizations of $L\times10^6$ renormalized lattice sites. We vary the strip width $L$ from $2$ up to $1024$. The variance $\sigma^2(E,L)$ of our data $\Gamma(E,L)$ follows the relation $\sigma^2(E,L)=\varsigma^2(L) \Gamma(E,L)$. Due to the constant number of realizations, the relative standard deviation $\varsigma(L)$ is independent of $E$ and scales as $\sqrt{L}$. Our numerical effort leads to $\varsigma(L)\approx 0.000052\sqrt{L}$. 

\subsection{Heuristic scaling} \label{subsec:heuristic}

\begin{figure*}
	\centering
	\includegraphics{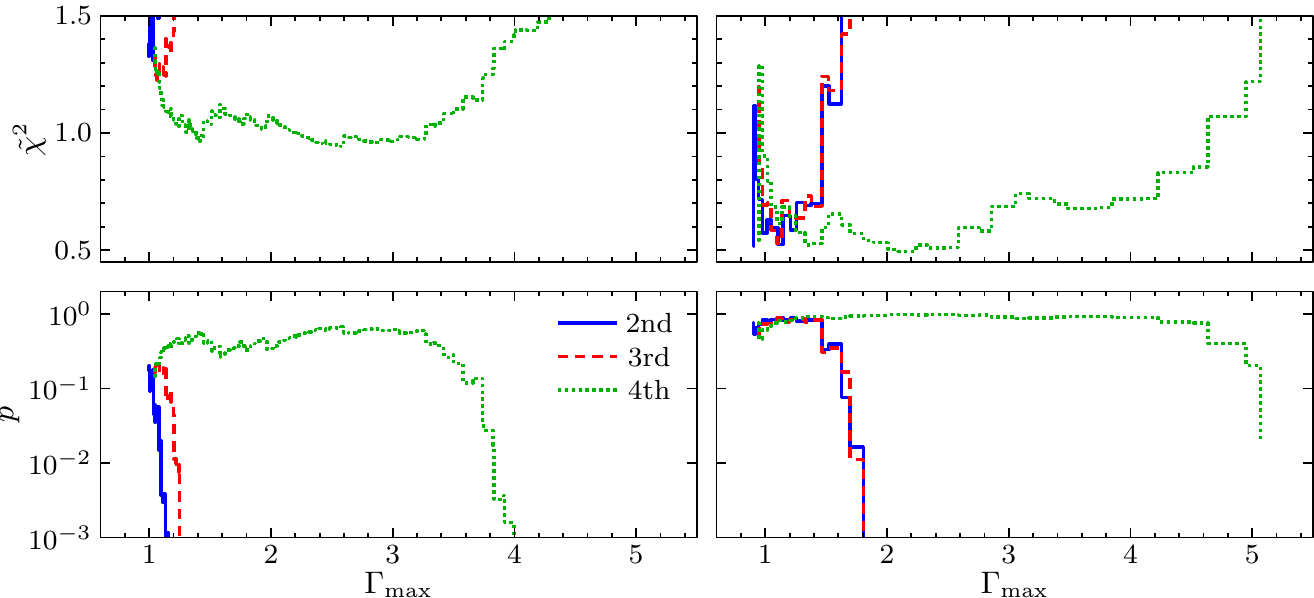}
	\caption{Analysis of the fit quality: Reduced mean-squared deviation $\tilde{\chi}^2$ and fit significance $p$, as functions of the data range (largest included $\Gamma$) for different orders of the fit polynomial (\ref{eq:polynomical}) for $L=32$ (left) and $L=1024$ (right).}
	\label{fig:fitrange}
\end{figure*}

In a first, preliminary step, we analyze the numerical data by means of a heuristic approach that discusses the scaling of different quantities independently. We start by fitting the energy dependence of the dimensionless Lyapunov exponent $\Gamma$ for each system size separately by a polynomial of up to fourth order in the distance from the minimum position $E_\mathrm{min}$,
\begin{align}
	\Gamma(E)=\Gamma^{(0)} + \Gamma^{(2)}(E-E_\mathrm{min})^2 + \Gamma^{(3)}(E-E_\mathrm{min})^3 + \Gamma^{(4)}(E-E_\mathrm{min})^4 \label{eq:polynomical}~.
\end{align}
To characterize the fit quality, we check the (reduced) mean-squared deviation $\tilde{\chi}^2$ and the significance $p$ of the fit for different orders of the fit polynomial (\ref{eq:polynomical}) and different data ranges. The latter are conveniently expressed in terms of the upper bound $\Gamma_\text{max}$ (all data points with $\Gamma(E_\text{min}) \leq \Gamma \leq\Gamma_\text{max}$ are included in the fit).
Figure \ref{fig:fitrange} shows examples of the mean-squared deviation $\tilde{\chi}^2$ and the significance of the fit $p$ for $L=32$ and $L=1024$.
$\tilde{\chi}^2$ and $p$ show a strong relation due to the (mostly) large number of free parameter. For strip width $L=32$, the data range for which a quadratic parabola leads to a good fit is extremely narrow and limited to the immediate vicinity of $E_\text{min}$ (almost invisible in Fig.\ \ref{fig:fitrange}). The inclusion of the cubic term widens the data range of reasonable fits somewhat. Considering 4th-order polynomials leads not only to a broad $\Gamma$ range for reasonable fits, but also to a higher overall fit quality than the lower-order polynomials. For larger strip width ($L=1024$), the behavior is similar, but quadratic and cubic polynomials provide descriptions of similar quality, indicating a less pronounced asymmetry in the data.
In the following, we therefore employ fourth-order polynomials to fit the $\Gamma(E)$ curves. The the values of $\tilde{\chi}^2$  and $p$ are used to identify the best data range for the further analysis. The resulting optimal fit ranges are highlighted in Fig.\;\ref{fig:transition} as solid lines.

\begin{figure*}
	\includegraphics{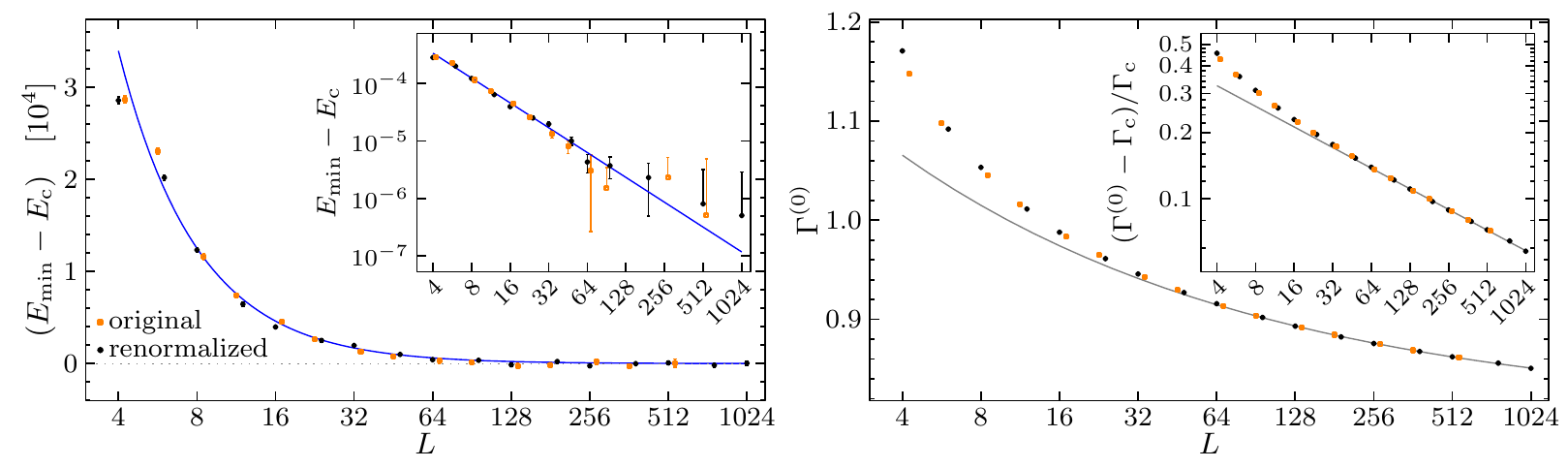}
	\caption{System-size dependence of the minimum of the $\Gamma(E)$ curves vs.\ strip width $L$, showing the minimum position $E_\text{min}(L)$ (left) and minimum value $\Gamma^{(0)}(L)$ (right), for calculations on both the original and the renormalized lattices. The system sizes of the original lattices are scaled by $1/\sqrt{2}$. The minimum position is plotted with respect to $E_\text{c}=3.4221475$, and the solid blue line shows the function $E_\text{min}(L)=E_\text{c}+a L^{-y'}$ with exponent $y'=1.45$. The value at the minimum, $\Gamma^{(0)}$, is plotted with respect to the asymptotic value $\Gamma_\mathrm{c}=0.804$, and the gray line shows the function $\Gamma^{(0)}(L)=\Gamma_\text{c}+a L^{-y}$ with exponent $y=0.31$. Insets show the same data as the main panels, but in a $\log$-$\log$ representation.}
	\label{fig:minimumscaling}
\end{figure*}

We now turn to the system-size dependence of the position $E_\text{min}$ and value $\Gamma^{(0)}=\Gamma(E_\text{min})$ of the minimum of the $\Gamma(E)$ curves. These data are presented in Fig.\ \ref{fig:minimumscaling}.
For comparison, the figure also shows data obtained for the original, unrenormalized lattice\;\cite{PusCSV19}. To compensate for the renormalization of the lattice constant, we scale the sizes of the unrenormalized lattices by $1/\sqrt{2}$, so that the largest strip width of 768 used in Ref.\;\cite{PusCSV19} corresponds to $L=768/\sqrt{2}\approx543$ in the renormalized lattice. The data for the original and renormalized lattices agree very well, as is expected because the Aoki renormalization group steps are an exact transformation of the Hamiltonian.

The data in Fig.\ \ref{fig:minimumscaling} demonstrate that the minimum position $E_\text{min}(L)$ converges quickly towards the critical energy $E_\mathrm{c}$. For sizes $L\gtrsim 96$, the system-size dependence of $E_\text{min}(L)$ becomes comparable or smaller than the statistical error of the data points. This rapid decay is not compatible with a marginal correction to scaling (at least for the studied system sizes); we thus use the power-law ansatz $E_\text{min}(L)=E_\text{c}+a L^{-y'}$. We exclude from the fit the values for extremely small $L\lesssim 4$, because they appear to suffer from additional finite-size effects. Fitting all data with $L\geq 8$, we obtain $E_\text{c}=3.4221469(10)$ and $y'=1.45(7)$ with a fair fit quality of $\tilde{\chi}^2=1.60$. The corresponding fit curve is shown in the left panel Fig.\;\ref{fig:minimumscaling}.
The quality increase when we neglect smaller systems. Using sizes $L\geq 16$ alters our estimates to $E_\text{c}=3.4221459(13)$ and $y'=1.18(13)$ with $\tilde{\chi}^2=1.23$, and for $L\geq 32$, we find $E_\text{c}=3.4221468(9)$ and $y'=1.8(3)$ with $\tilde{\chi}^2=0.87$. The inset of  Fig.\;\ref{fig:minimumscaling} illustrates the difficulties in getting a reliable estimates for $y'$, as the systematic $L$ dependence becomes smaller than the statistical uncertainties for larger $L$.

In contrast, the dimensionless Lyapunov exponent value $\Gamma^{(0)}$ at the minimum features significant finite-size corrections even for the largest
studied system sizes. We first model these corrections using the ansatz $\Gamma^{(0)}(L)=\Gamma_\text{c}+a L^{-y}$.  For strip width $L\geq 128$, we find the estimates $\Gamma_\text{c}=0.813(5)$ and $y=0.36(3)$ with $\tilde{\chi}^2=0.81$. A similarly good fit (with compatible results) can be obtained for $L\geq 64$, but if smaller $L$ are included, a single power-law correction is not sufficient for a high-quality fit. Including an additional term with exponent $2y$ does not improve the fit quality.
We note that the description of $\Gamma^{(0)}$ in terms of a single power-law correction breaks down at system sizes where $E_\text{min}$ deviates from an approximately constant value and the correction term with exponent $y'$ becomes important. We therefore also consider the ansatz $\Gamma^{(0)}(L)=\Gamma_\text{c}+a L^{-y}+b L^{-y'}$ which contains two correction terms. Fixing $y'$ at the rough estimate $y'=1.3$ (based the $E_\text{min}$ results above), we obtain high-quality fits ($\tilde{\chi}^2\approx 0.6$) for all $L$-ranges from $L\ge 8$ to  $L \ge 64$.  They yield consistent parameter estimates that can be summarized as $\Gamma_\text{c}=0.805(3)$ and $y=0.31(3)$. If we let both $y$ and $y'$ float, the fits are stable and of equally good quality for lower fit range boundaries between $L=6$ and $L=24$. For example, the fit range $L\geq 8$ yields $\Gamma_\text{c}=0.807(2)$, $y=0.316(9)$, and $y'=1.36(5)$. The great consistency of the parameter estimates supports the scenario of two independent irrelevant exponents $y$ and $y'$.

However, due to the slow approach of $\Gamma^{(0)}(L)$ towards its asymptotic value, we cannot exclude marginal corrections. Using the ansatz $\Gamma^{(0)}(L)=\Gamma_\text{c}(1+a/(b+\log L))$, we obtain good fits ($\tilde{\chi}^2\lesssim 1$) for all size ranges above $L = 16$, yielding  $\Gamma_\text{c}=0.7381(9)$ and $b=0.63(2)$. Note that the constant $b$ defines a characteristic length $L_0=\exp(-b)\approx 0.53$ (in multiples of the renormalized lattice constant). If we add an additional power-law term with exponent $y'$, the $L$-range for good fits increases to $L\geq8$, and we find  $\Gamma_\text{c}=0.742(4)$ and $b=0.28(11)$ with $y'=0.96(15)$.

In order to determine the localization length critical exponent $\nu$, we analyze the curvature $\Gamma^{(2)}$ of the $\Gamma(E)$ curves at their minimums. We start with an ansatz that contains one power-law correction to scaling, $\Gamma^{(2)}(L)= L^{2/\nu}(1+ a L^{-y})$. Unfortunately, fits in which both $\nu$ and $y$ float turn out to be numerically unstable. To test whether the new numerical data are compatible with our previous results \cite{PusCSV19}, we perform fits with $\nu$ fixed at $\nu=2.6$. As in Ref.\ \cite{PusCSV19}, this description leads to good fits for large systems only. From fits over the size ranges $L\geq128$ and $L\geq64$ we obtain, respectively, $y=0.36(9)$ and $y=0.299(15)$. Alternatively, we can perform fits in which $\nu$ and $y$ are connected by the proposed holomorphic relation\;\cite{LutR07}, $y=1/\nu$. This yields $y=0.3859(17)$ or $\nu=2.591(12)$ for the fit range $L\geq 128$ with $\tilde{\chi}^2=0.77$. For $L\geq64$, the results read $y=0.3893(9)$ [$\nu=2.568(6)$] with $\tilde{\chi}^2=1.09$. For high-quality fits over a wider system-size range, we need to add further corrections-to-scaling terms. For $L\geq16$, we are able to obtain reasonable results if we set $y'\approx 1$. However, the accuracy of our $\Gamma^{(2)}$ data is not sufficient to separate different power-law contributions reliably. We also considered marginal corrections to scaling combined with a fixed value $\nu=2.6$. This model describes our data well for $L\gtrsim96$, but the estimates of $b$ have large errors.

\subsection{Sophisticated scaling}

In the previous section we studied the scaling behavior of different quantities separately. We now focus on a combined description of the entire Lyapunov exponent data set $\Gamma(E,L)$ in the vicinity of $E_\text{c}$ for a range of system sizes $L$. To this end, we employ the sophisticated scaling approach introduced in Sec.\;\ref{sec:SSA}. Within this approach, the same exponents must describe the corrections to scaling for all quantities. This puts additional constraints on their values and is thus expected to stabilize the fits.

We present in Tab.\;\ref{tab:SSA_critparam} the parameters and results of fits of our data with the scaling function defined in eqs.\ (\ref{eq:fss})  and (\ref{eq:complex_scaling_field}) for various expansion orders and data ranges.
\begin{table}
	\caption{Critical parameter estimates using the sophisticated scaling approach of Sec.\;\ref{sec:SSA} as functions of system size range and fit expansion order. We include all systems sizes $L\geq L_\text{min}$. The expansion order is expressed as the quintuple ($n_\text{i}, n'_\text{i}, n_\text{r}, m_\text{r}, m_\text{i}$). The first column includes labels for fits we address in the main text. Bracketed numbers denote the standard deviation of the associate value. We mark fits of 'reduced' quality, $p\leq 0.1$, by italic $p$ numbers.}
	\label{tab:SSA_critparam}	
	\setlength\tabcolsep{4 pt}
	\centering
	\begin{tabular}{c|l|l|l|l|l|c|c|r|r|r|r}
		\#  &$~~~~\Gamma_\mathrm{c}$& $~~~~~~E_\mathrm{c}$& $~~~~~\nu$ & $~~~~~y$ & $~~~~y'$ & $L_\text{min}$ & order & $N_\text{D}~$ &  $N_\text{P}$ & $\tilde{\chi}^2~~~$ & $p~~~$ \\
		\hline
		& 0.803(2)  & 3.4221482(6) & 2.618(7)  & 0.298(9)  & 1.20(10)  & 6 & 1\,1\,3\,3\,2 & 1135 & 24 & 1.0026 & 0.4700\\
		& 0.811(1)  & 3.4221490(6) & 2.591(3)  & 0.334(4)  & 1.60(2)   & 6 & 1\,1\,3\,3\,1 & 1135 & 23 & 1.0560 &\textcolor{red}{\it 0.0950}\\ 
		d & 0.808(2)  & 3.4221483(5) & 2.602(4)  & 0.321(6)  & 1.52(3)   & 8 & 1\,1\,3\,3\,1 & 1047 & 23 & 0.9849 & 0.6288\\
		c & 0.809(11) & 3.4221485(4) & 2.63(3)   & 0.32(9)   & 1.0(3)    & 8 & 1\,2\,2\,3\,2 & 1047 & 26 & 1.0177 & 0.3400\\ 
		& 0.807(5)  & 3.4221493(4) & 2.611(9)  & 0.31(3)   & 1.53(3)   & 8 & 1\,2\,2\,3\,1 & 1047 & 25 & 1.0251 & 0.2818\\ 
		& 0.808(4)  & 3.4221482(6) & 2.609(6)  & 0.317(18) & 1.32(7)   & 16 & 1\,1\,3\,3\,1 & 871  & 23 & 0.9680 & 0.7421\\
		& 0.804(4)  & 3.4221485(5) & 2.609(13) & 0.300(18) & 1.26(7)   & 16 & 1\,1\,2\,3\,1 & 871  & 19 & 1.0263 &  0.2897\\
		& 0.806(6)  & 3.4221480(6) & 2.612(12) & 0.30(3)   & 1.55(19)  & 32 & 1\,1\,3\,3\,1 & 695  & 23 & 0.9268 & 0.9127\\
		& 0.805(6)  & 3.4221483(5) & 2.612(13) & 0.30(3)   & 1.52(19)  & 32 & 1\,1\,2\,3\,1 & 695  & 19 & 0.9837 & 0.6114\\
		a & 0.808(3)  & 3.4221484(6) & 2.67(13)  & 0.3(2)    & --        & 64 & 2\,0\,3\,3\,0 & 519  & 17 & 0.9758 & 0.6426\\
		b & 0.807(3)  & 3.4221484(7) & 2.563(5)  & 0.328(7)  & --        & 64 & 1\,0\,3\,3\,0 & 519  & 13 & 1.2014 & \textcolor{red}{\it  0.0012}\\ 
		& 0.805(3)  & 3.4221482(7) & 2.580(8)  & 0.313(11) & --        & 96 & 1\,0\,3\,3\,0 & 431  & 13 & 0.9873 & 0.5641\\
		& 0.805(3)  & 3.4221483(6) & 2.581(6)  & 0.316(11) & --        & 96 & 1\,0\,2\,3\,0 & 431  & 11 & 1.0756 & 0.1374\\
		& 0.805(6)  & 3.4221479(8) & 2.590(11) & 0.314(16) & --        & 128 & 1\,0\,3\,3\,0 & 351  & 13 & 0.9128 & 0.8733\\
		& 0.800(5)  & 3.4221489(6) & 2.560(5)  & 0.288(17) & --        & 128 & 1\,0\,3\,2\,0 & 351  & 12 & 0.9691 & 0.6483\\
		& 0.804(4)  & 3.4221475(6) & 2.598(8)  & 0.312(16) & --        & 128 & 1\,0\,2\,3\,0 & 351  & 11 & 1.0282 & 0.3483\\
		& 0.77(3)   & 3.4221488(9) & 2.64(4)   & 0.20(7)   & --        & 256 & 1\,0\,3\,3\,0 & 219  & 13 & 0.9507 & 0.6829\\
		& 0.77(3)   & 3.4221483(8) & 2.62(3)   & 0.22(6)   & --        & 256 & 1\,0\,2\,2\,0 & 219  & 10 & 1.0157 & 0.4238\\
		& 0.82(3)   & 3.4221486(11)& 2.56(6)   & 0.4(4)    & --        & 512 & 1\,0\,2\,2\,0 & 114  & 10 & 0.9254 & 0.6934\\
	\end{tabular}
\end{table}

Similar to the findings in the previous section, the data for system sizes $L\geq 64$ can be well fitted by including only one
(leading) irrelevant exponent $y$. For $L\geq 96$, the simplest possible irrelevant correction, $n_\text{i}=1$, is sufficient to obtain
high-quality fits. If $L=64$ is included, the quality of a fit with $n_\text{i}=1$ suffers but increasing the irrelevant expansion
order $n_\text{i}$ helps to raise the fit's quality (see rows a and b in the table).
When extending the fits to smaller system sizes, raising $n_\text{i}$ is insufficient to obtain good fits. Instead, a subleading term must be included, governed by the exponent $y'$ (but the irrelevant expansion orders can remain at $n_\text{i}=1$ and $n'_\text{i}=1$). For very small systems (system-size range  $L\geq8$), we increase the relevant expansion order $n_\text{r}$ which leads to a better fit and uses fewer fit parameter than raising the irrelevant expansion order $n'_\text{i}=2$ (see rows c and d in the table). However, even for larger systems without evident subleading contributions, the increase of $n_\text{r}$ from 2 to  3  helps improving the fit quality. It leads to only minor changes in the critical parameters, emphasizing the robustness of the fits.

The critical parameters resulting from different fits within the sophisticated scaling approach agree very well with each other, including the fits with and without subleading corrections to scaling. We summarize them as follows. The system undergoes a phase transition at $E_\text{c}=3.4221485(10)$, where the (asymptotic) critical Lyapunov exponent takes the  value $\Gamma_\text{c}=0.806(6)$. The critical fixed point is characterized by the relevant exponent $\nu=2.60(2)$ and the leading irrelevant exponent $y=0.31(3)$. We also identify a subleading irrelevant exponent $y'$ significantly larger than $y$. In contrast to other quantities, its estimates from different fits deviate more strongly from each other
than their individual statistical uncertainties. We therefore summarize them as $y'=1.4(2)$. So far, the discussion has focused mainly on the robustness of the fits w.r.t. different system-size fit ranges and expansion orders of the scaling function. We have also verified that the critical parameters are robust against changes in the size of the energy (or $\Gamma$) range included in the fits.

In addition, we also consider fits that combine a leading marginal (logarithmic) correction to scaling, expanded in terms of $1/(b+\log L)$, with a subleading power-law correction. For a better comparison, we employ the same expansion orders as used for the power-law fits with $y$ and $y'$. If we restrict the fits to large system sizes, the subleading correction is not required, and we obtain fits of similar quality to the power-law scenario above, yielding $\Gamma_\text{c}=0.73(1)$, $\nu=2.57(2)$, and $b=0.8(4)$. When the fit range is extended to include smaller sizes that require the addition of the subleading term, the fits become unstable. Estimates for $b$ cover very wide ranges and depend strongly on the data range and fit expansion orders. We hence consider this scaling scenario to be less appropriate.

\section{Conclusion} \label{sec:summary}
To summarize, we have presented an improved method to simulate localization problems for tight-binding Hamiltonians. It employs exact renormalization group steps\;\cite{Aok80} to construct a renormalized tight-binding lattice with half the number of lattice sites. The renormalized Hamiltonian is then studied in a quasi-one-dimensional geometry by means of the recursive Green's function technique. We have applied this method to investigate the integer quantum Hall transition on simple square lattices. This novel approach has accelerated our computations by factor of four in comparison to our previous analysis \cite{PusCSV19} which applied the recursive Green's function technique to the original lattice. We have used the performance gain to improve the accuracy of our data and to double the effective size of our systems. This substantial improvement helps us to analyze the IQH transition which suffers from strong finite-size effects whose form remains a puzzle in the quantum Hall field.

Our numerical data reveal the presence of a subleading correction to scaling that affects smaller systems of sizes $L\lesssim 64$. This correction decays rapidly with $L$, the corresponding subleading irrelevant exponent $y'$ is significantly larger than the estimate for the leading irrelevant exponent $y$. Supported by a heuristic scaling approach, we have custom-tailored a sophisticated scaling function that includes two independent irrelevant exponents, $y$ and $y'$, in addition to the relevant exponent $\nu$. This scaling function is able to describe consistently
the scaling behavior of our numerical data for the dimensionless Lyapunov exponent $\Gamma$  for linear system sizes covering two orders of magnitude.
The fits yield $\Gamma_c=0.806(6)$, $\nu=2.60(2)$, $y=0.31(3)$, and $y'=1.4(2)$. These results agree with and improve upon  our previous investigation\;\cite{PusCSV19}, where the main analysis was based only on systems for which the subleading corrections could be neglected.
In contrast to recent suggestions, we do not see any evidence to prefer marginal corrections to scaling over power-law corrections.

Some properties of the subleading correction to scaling remain unclear. Our analysis shows that the (effective) critical energy, as defined via the minimum of the $\Gamma(E)$ curve, features a rapidly decaying finite-size correction governed by $y'$. Is $y'$ an independent exponent or a multiple of $y$ (stemming from a higher order term in the expansion)? Our numerical estimates would be compatible with $y'=4y$ or $y'=5y$. If $y'$ is independent, is it universal? Moreover, one may ask how the two corrections to scaling appear in various observables and, in particular, why the leading irrelevant exponent appears to be missing in the $L$-dependence of the effective critical energy. Answering these questions requires simulations of significantly larger systems and remains a task for the future.

\subsection*{Acknowledgment}
This work was supported by the NSF under Grant Nos.\ DMR-1506152, DMR-1828489, PHY-1125915, PHY-1607611, and OAC-1919789.  M.P. acknowledges the support from German Research Foundation (DFG) through the Collaborative Research Center, Project ID 314695032 SFB 1277 (projects A03, B01). We thank Jo\~ao C. Getelina and Jos\'e A. Hoyos for helpful discussions regarding the renormalization group steps.

\end{document}